# Non-Driving-Related Tasks Influencing Drivers' Takeover Time: A Meta-Analysis

Yan Zhang

Department of Systems Engineering and Engineering Management

City University of Hong Kong

Abstract

Before the era of fully automated vehicles, human is consistently an indispensable part of the driving system. Various studies have investigated drivers' cooperation with the vehicle under different conditions. In this article, we analyzed how non-driving-related tasks (NDRT) influence takeover time (TOT) by conducting a meta-analysis on 37 related papers. NDRTs were transcoded into combinations of five basic dimensions to unify and demonstrate their effects on drivers. In order to interpret experimental data comprehensively, we implemented three methods. A synthetical analysis was conducted to compare the effect size between each study and subgroup. Studies with eligible control groups have been examined by the two-group analysis, followed by moderator analysis on seven variables. The results from the two-group analysis showed that both visual-mental-motoric and visual-mental tasks have significant negative effects on the takeover time and the previous type had a larger effect than the latter one. Moreover, the subgroup comparison and meta-regression in the meta-analysis part revealed the correlation between moderators and the effect size, in which the Driving Experience and the Automation Level affected the relation between NDRT and TOT. The findings of this paper can contribute to the improvement and new directions for further scientific research and engineering design.

Keywords: Automated driving, Takeover time, Non-driving-related task, Meta-analysis

1. Introduction

As automated driving technology is developing rapidly, it is evident and inexorable that self-driving vehicles will become mainstream. In the ideal state, the automated driving system (ADS) can save time and manpower for individuals and factories under safe circumstances. However, the development of any technology requires stages. The American Society of Automation Engineers (SAE) has defined automated driving as 6 levels, which are No Automation (L0), Driver Assistance (L1), Partial Automation (L2), Conditional Automation (L3), High Automation (L4), and Full Automation (L5). According to the definition by the Federal Highway Research Institute (BASt) (Gasser & Westhoff, 2012), except for full automation, all other levels of systems require human intervention when facing emergencies or confusing conditions. Therefore, it is consequential to study the interaction between drivers and vehicles. In this study, the influence of non-driving-related tasks (NDRT) on drivers' takeover times (TOT) was the focus of discussion.

1.1 The necessity of drivers' existence in non-fully automated vehicles

According to SAE, a level 0 automated system gains no control of the vehicle but is able to issue warnings to drivers. Driver assistance (L1) and partial automation (L2) have been maturely used in control systems for years. The level 1 autopilot system controls the speed or steering, while the level 2 controls the two main functions at the same time. Before conditional automation, drivers are still required to pay attention to the road condition all the time. After reaching level 3 and level 4, drivers are allowed to distract their sights or minds, whereas they need to take back the control when receiving takeover requests from the car.

In reality, the road conditions are rather complex and empirical. It is conceivable for the system to receive indistinct environment information. For instance, on May 7th, 2016, a fatal collision was caused by a Tesla Model S as its driver was not able to retrieve the control when the system failed to respond to a tractor-trailer crossing the intersection on a highway. The accident analysis concluded there was no flaw in the design and performance of the system whereas human error was the one to be blamed (National Highway Traffic Safety Administration). Therefore, the prompt intervention of manual control is essential.

Even though the level 3 automation system has been implemented by some pioneer companies and level 4 automation was qualified in trial runs (Martínez, 2021), high automation still cannot get legal support when implemented on public roads. Based on the present circumstances, the application of full automation requires decades (Kyriakidis, Happee, & de Winter, 2015; Petermeijer, Bazilinskyy, Bengler, & De Winter, 2017), which means humans are still an important part of the driving system.

1.2 Interaction between driver and vehicle

Switching between automated and manual driving is unavoidable when using any level of non-fully automated vehicles. Compared with a controllable switch from manual driving to autopilot, the unpredictable and urgent takeover request (TOR) from autopilot to manual driving is more vital. The takeover response takes time and an accurate prediction of the time required to resume control is critical to the design of systems (Beller, Heesen, & Vollrath, 2013). Therefore, the measurement and prediction of drivers' takeover time (TOT) are significant.

The interaction exists in any human-machine system. Several published papers showed that the automated system may cause negative effects on human performance during the driving process. Stanton and Marsden (1996) indicated driving skills can fade away if automation took dominance. When facing an emergency, the driver can become unqualified to cope with the situation. Another study from Endsley (1999) made the point that the combination of the human and the machine performed worse than either machine or human as separate, because of decision making. When the computer provides options, the operator can be confused and distracted, especially when the decision from the machine has an inconsistency with the drivers' judgment. A driver with less confidence in driving skills may end with an overreliance on the automated vehicle (Parasuraman & Riley, 1997), which may cause panic when a takeover request pops out. However, the current automated vehicle tests are mainly concentrated on the driving systems rather than the human aspect and the interaction in-between (Xing, Lv, Cao, and Hang, 2021).

1.3 Non-driving-related task

One thing that accompanies the application of automation is the non-driving-related task (NDRT). Drivers may utilize the benefit of an automated system and distract their attention, which is understandable because it is the goal of autopilot. However, without full automation, this kind of distraction will lead to the out-of-the-loop (OOTL) problem. The OOTL problem has been defined as a common automation negative effect (Billings, 1991; Wiener & Curry, 1980). When immersing in the OOTL state, it has been proved that operators have a weakened ability to detect problems and take back manual controls (Endsley & Kiris, 1995). When a TOR is signaled, there is a high probability that the driver is performing a secondary task and in an OOTL state.

In order to measure the takeover time under the out-of-the-loop situation, non-driving-related tasks are designed and used in research. Performing secondary tasks can help participants simulate the OOTL situation. Louw and Merat (2017) have noticed that when drivers were performing a secondary task, they were more likely to lose attention to the control system, even when the road was invisible during rough weather. By conducting a comprehensive literature review, the existing NDRTs used in experiments were classified into two types. The first type of NDRT was designed based on reality. Researchers considered what a driver would do when driving an automated vehicle. Tasks under this category mostly were smartphone interaction, watching a video, playing games by using the in-car tablet, etc. (Yoon, Kim, & Ji, 2019; Cohen-Lazry, Katzman, Borowsky, & Oron-Gilad, 2019; Wintersberger, Riener, Schartmüller, Frison, & Weigl, 2018) Another type of NDRT were designed purely by the dimensions of distraction, which include visual distraction, auditory distraction, mental distraction, and motoric distraction. (Naujoks, Purucker, Wiedemann & Marberge, 2019) For example, when an auditory distraction was expected in the experiment, Naujoks decided to use audiobooks as the secondary task.

## 1.4 Aims and Objectives

Performing a non-driving-related task during automated driving is considerable so that enter the OOTL state is an unavoidable issue before the era of full automation. Researchers have made great efforts on exploring the relations between TOTs and NDRTs. In this paper, the meta-analysis was used to composite relevant results from sufficient studies to generate comprehensive conclusions. The ultimate goal of this study was to investigate how the non-driving-related tasks influence drivers' takeover time. The results can provide a verifiable reference for refining the ADS.

## 2. Methodology

### 2.1 Data selection

270 results had been found from Google Scholar by searching with specific qualifications. Keywords that described the takeover reaction from automated driving to manual driving and synonyms of non-driving-related-tasks had been used, together with an emphasis on "experiment" and "driver" to narrow the scope. All patents and non-English results were excluded. The results were conducted by the following statement: ("take over" OR takeover) AND "automated driving" AND "manual driving" AND ("reaction time" OR "takeover time") AND "automated vehicle" AND experiment AND driver AND ("secondary tasks" OR NDRT OR NDT).

To ensure the validity of the analysis, all literature has been screened by the following criteria:

1. The study had to be experimental. No review or theoretical paper was considered.

2. The experiment had to be related to NDRTs and the type of NDRTs must be specified. The existence of control groups with no NDRT was not a necessity.

3. The result had to contain takeover times. The measurement and definition of the takeover time should be stated in the method section.

4. The study had to declare the means and standard deviations of takeover times, or the means and standard deviations can be calculated by the presented data. The sample size of each group should also be available.

After a bare-bones screening of the topics and abstracts, 7 duplications and 95 non-related papers have been removed. 168 articles were assessed exhaustively, in which 41 non-experimental studies, 16 non NDRT related studies, 8 studies with unclarified NDRT modality, 45 resulted without takeover time, and 21 articles with no obtainable mean or standard deviation have been eliminated. Eventually, 37 eligible studies were included in this paper. Figure 1 illustrates the selection procedure.

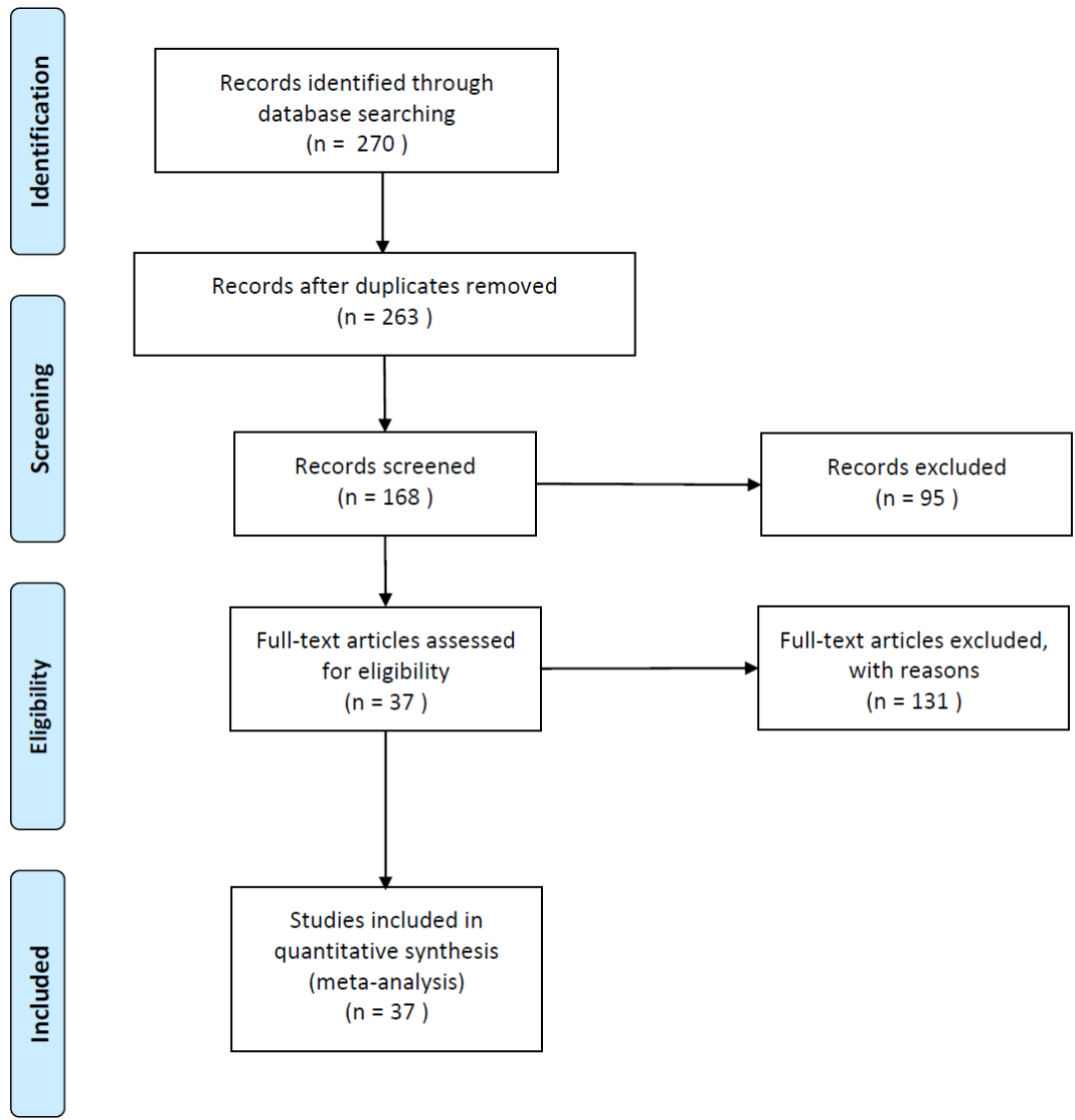

Figure 1. Selection criteria

## 2.2 Coding

The modalities of NDRT are composed of 5 basic dimensions. N for No-NDRT, Vi for visual tasks, Me for mental tasks, Au for auditory tasks, and Mo for motoric tasks (Table 1). For the literature, most secondary tasks used by researchers were not purely one-dimensional. As explained in the introduction, some experiments may not state the modality explicitly but define the exact task. As a matter of fact, by the nature of the actual driving condition, it is unusual that the secondary task only involves one dimension. The most popular used tasks were reading an article and watching a video. As a result, dimensions that a task involves have been distinguished by which awareness and mobility it occupies. For example, a game task was transcoded as a combination of Vi, Me, and Mo because it required the driver to watch, think, and tap. In this paper, all the tasks have been transcoded into combinations of basic dimensions. The records of all the combinations are shown in Table 2. The description of the original tasks of each paper can be found in Appendix A. The valid takeover time measurement has been defined as the duration between signaling the takeover request and the drivers' first action. To unify all the experiments from different articles, TOTs were documented in milliseconds.

When reviewing the literature, other than NDRT, seven variables were found frequently discussed because of their influence on takeover time during experiments, which were the age, sex ratio, driving experience of the participants, automation level of the vehicle, and the type of takeover request (visual, auditory, and vibrotactile takeover request). In order to diminish the interference of these moderators in this meta-analysis, the moderator analysis was conducted to analyze their impacts. The detailed descriptions are shown in Table 3.

Table 1. Five basic dimensions for NDRT

| Dimension of NDRT | Abbreviation |
|---|---|
| No NDRT | N |
| Visual | Vi |
| Auditory | Au |
| Mental | Me |
| Motoric | Mo |

Table 2. Statistics of combinations

| Modality | Number |
|---|---|
| N | 15 |
| Au | 2 |
| Mo | 1 |
| Au Me | 2 |
| Vi Me | 9 |
| Vi Mo | 1 |
| Vi Au Me | 5 |
| Vi Au Mo | 1 |
| Vi Me Mo | 31 |

Table 3. Descriptions of moderators

| Moderator | Coding | Detail |
| --- | --- | --- |
| Age | Years | Mean age of participants |
| Sex Ratio | Male: Female | Sex ratio of participants |
| Driving Experience | Years | Participants' driving experience |
| Automation Level | 2 = Level 2 | Automation level of the apparatus |
| | 3 = Level 3 | |
| | 4 = Level 4 | |
| VTOR | 0 = No | Presence of the visual takeover request |
| | 1 = Yes | |
| ATOR | 0 = No | Presence of the auditory takeover request |
| | 1 = Yes | |
| ViTOR | 0 = No | Presence of the vibrotactile takeover request |
| | 1 = Yes | |

2.3 Documentation

Appendix A is the complete documentation of all eligible studies. Secondary tasks have been transcoded into a unified format. It is hard to ignore that not every study has a control group as the data may be extracted from a study that did not focus on comparing the impact of the existence of the secondary tasks. To be more specific, the main object can be finding out the effect between different kinds of secondary tasks. (Naujoks et al., 2019) Therefore, the control groups, which were the N groups in this article, may not be available. All the necessary data, including sample size, mean, and standard deviation, were also documented.

2.4 Meta-analysis

The calculation was achieved by Comprehensive Meta-analysis Software. The study was based on the random effect model given that there was a huge difference between experiment methods and participants.

As stated in the previous section, discovering the effect of NDRT on TOT was not the primary purpose for every reviewed article. The comparison of two groups in meta-analysis requires both experimental groups and control groups. Studies that were paired with control groups were eligible for two-group analysis. To maximize the usage of collected data, studies with only experimental groups were used to perform synthetical analysis, which included a one-group analysis and comparison of subgroups. The one-group analysis in the meta-analysis only requires experimental groups. By classifying subgroups, the takeover time for different modalities can be clearly compared.

In the two-group analysis, Hedge's g, a measure of effect size, and its 95% CI were used. The negative effect was defined when the experimental group has a greater result, i.e., longer reaction time, than the control group. The funnel plot of precision by Hedge's g was provided to estimate the publication bias. $I^2$ values were used to investigate the heterogeneity of studies. When $I^2$ is larger than 50%, it indicates heterogeneity (Higgins, Thompson, Deeks, & Altman, 2003). Furthermore, there were seven variables influencing the TOT under different NDRT conditions. In order to identify their impacts, the two-group analysis was followed by a moderator analysis.

2.4.1 Synthetical analysis

The aim of the synthetical analysis was to reveal the statistical result among all experimental groups. Firstly, regardless of the existence of the control group, a one-group analysis was implemented on all modalities of NDRT. The overall random effect was analyzed. Secondly, the studies were grouped into subgroups to compare between modalities. Forest plots helped to visualize the effect. Because of the lack of control groups, means and 95% confidence intervals were the effect measure.

### 2.4.2 Two-group analysis

In this analysis, only studies with both control groups and experimental groups were considered. Therefore, further screening has been made in advance. Table 4 is a summarization, in which eligible studies were categorized by the modality of NDRT.

Table 4. Eligible studies for the two-group analysis

| Modality | Number |
|----------|--------|
| N | 15 |
| Au | 2 |
| Mo | 1 |
| Au Me | 2 |
| Vi Me | 6 |
| Vi Mo | 1 |
| Vi Au Me | 3 |
| Vi Me Mo | 13 |

Not all types of NDRT were popular among experiment designs. Some modalities, like motoric tasks, only appeared once. A meta-analysis is recommended to be performed when at least four studies are available (Fu et al., 2011). In another word, Vi Me (6) and Vi Me Mo (13) conditions were considered by this part.

### 2.4.3 Moderator analysis

The moderator analysis was conducted based on the table of possible moderators (Table 3). When screening the literature, frequently mentioned variables and participants' demographics were recorded and summarized, even though some of them were left blank because of the lack of data. To determine whether the relation between NDRT and takeover time was affected by a third variable, the control groups (N) are needed. Therefore, the moderator analysis was performed based on the two-group analysis.

## 3. Result

### 3.1 Study characteristics

Thirty-seven eligible studies gathered 2474 samples. The mean takeover times across studies with and without NDRTs were 3464ms and 3019ms, respectively. The takeover time of the experimental group was ranged from 9540ms to 750ms. By comparison, the control group was more concentrated, which was ranged from 7840ms to 1360ms.

### 3.2 Synthetical analysis

The forest plot (Figure 2) gathered all experimental groups of the reviewed studies and clustered them into subgroups. The TOT of most studies was smaller than 4500ms. The overall random effect was 3452.1ms. Figure 3 showed the comparison of subgroups. Au, Au Me, Mo, and Vi Au Me groups had similar mean TOT. Vi Me groups had the greatest value (5178ms) and followed by Vi Me Mo groups (3349.2ms). Vi Me Mo groups had the largest sample size and Vi Me groups had the largest standard error. Meanwhile, Vi Au Mo and Vi Mo groups had extremely small means and standard errors. As both of them only included one study, these values may have no statistical significance.

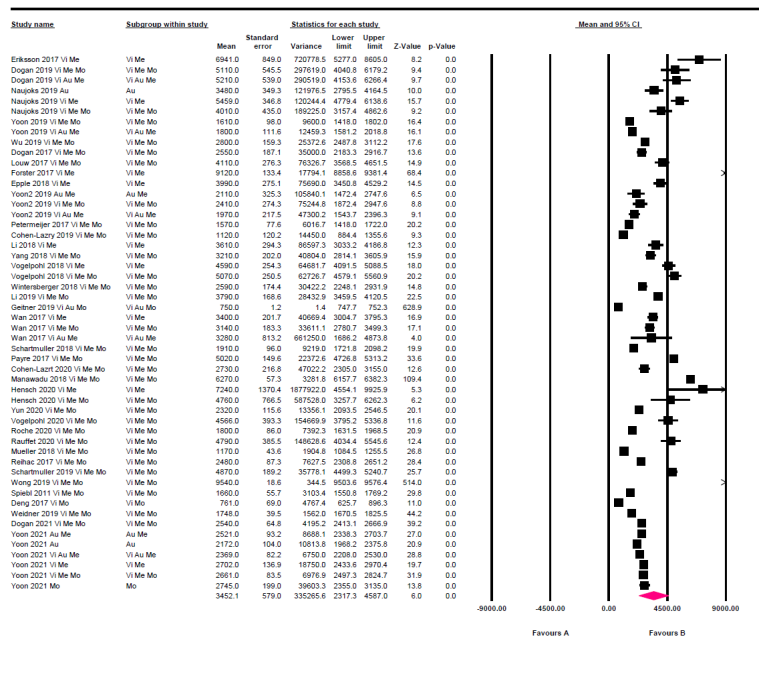

**Meta Analysis**

Figure 2. Statistic result of all experiment groups

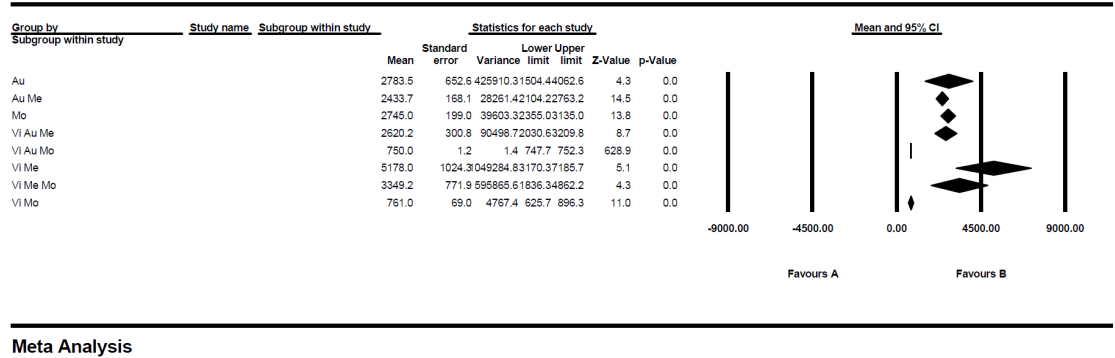

Figure 3. Comparing results of each subgroup

## 3.3 Two-group analysis

### 3.3.1 Vi Me Mo

When doing the Vi Me Mo two-group analysis, 13 studies were selected based on the criteria that those studies should contain both the control groups and the Vi Me Mo experimental groups.

Figure 4 shows Hedges's g and 95% CI of the random-effect model. It was distinct that all the g values were negative, and the mean effect size was -0.827, which indicated the experimental groups had a longer reaction time than the control groups. As the absolute value of the mean Hedges's g was larger than 0.8, by following the principle of Cohen (1988), it can be considered as a large effect. Most of the studies had an effect size within the range of -1 and 0. Even though some of the studies had a 95% confidence interval larger than one, the average CI was relatively small, which was ranged from -0.951 to -0.704. Ten out of thirteen studies had a p-value lower than 0.05. The overall p-value proved that the visual, mental, and motoric tasks had a significant negative effect on drivers' takeover time.

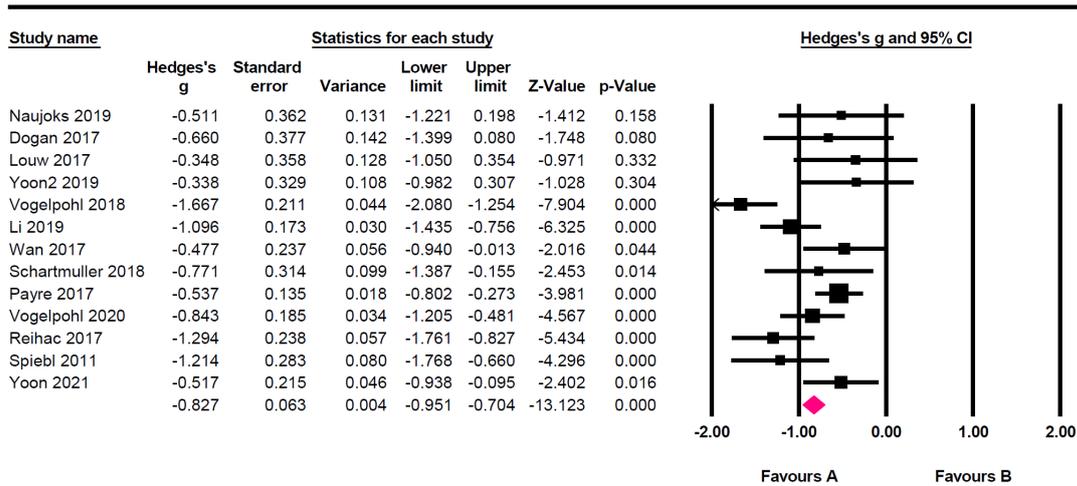

Figure 4. Meta-analysis result for Vi Me Mo

The analysis showed the fact that $I^2$ equaled 68.08 (P<0.001), indicating systematic heterogeneity between studies. The Funnel plot of precision showed no sign of publication bias and no data point was imputed (Figure 5).

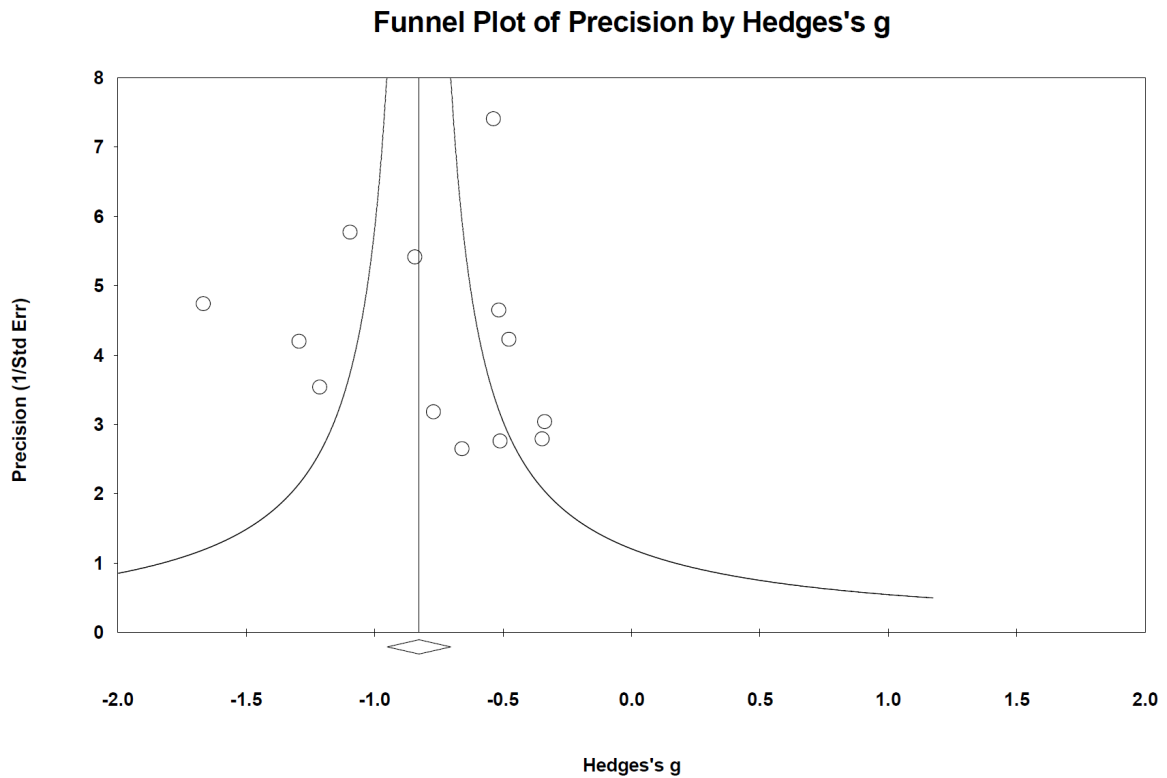

Figure 5. Funnel plot for Vi Me Mo

### 3.3.2 Vi Me

Six Vi Me studies were eligible for two-group analysis. The average takeover time for the experimental group was 5369ms, which was 1318ms greater than the control group.

The Hedges's g was ranging from -0.189 to -1.355 and the average value was -0.773 (p<0.001), which indicated a large negative effect on the takeover time. The average CI fell in the range of -0.393 to -1.153. Four out of six studies had a significant result (p<0.05). (Figure 6)

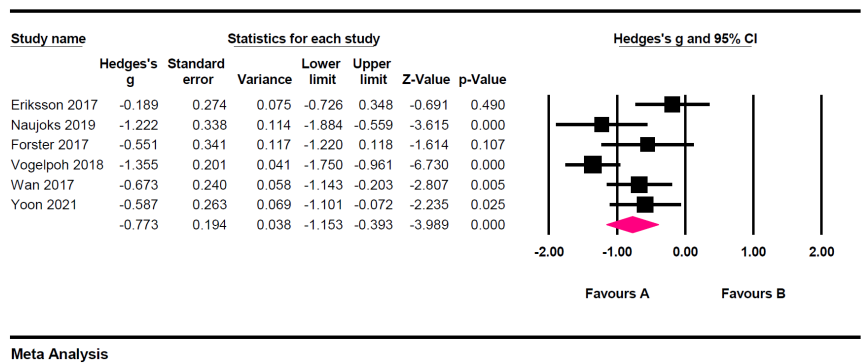

Figure 6. Meta-analysis result of Vi Me

The $I^2$ (67.89, p<0.05) of the heterogeneity test indicated there were unrevealed effect of moderators between studies. The funnel plot (Figure 7) showed a relatively symmetric distribution of the data point and no publication bias was found.

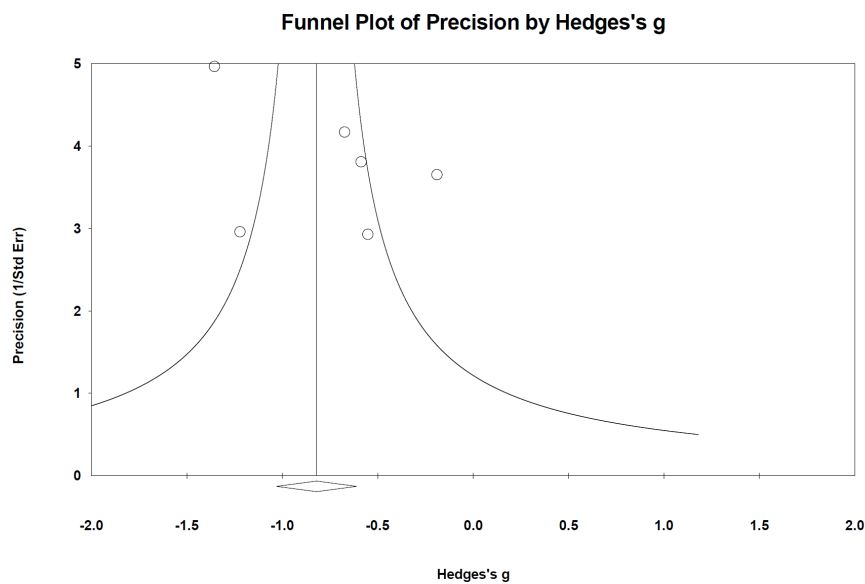

Figure 7. Funnel plot for Vi Me

3.4 Moderator analysis

3.4.1 Vi Me Mo

For the Vi Me Mo group, the calculated $I^2$ was 68.08 (P<0.001). Figure 8 showed the forest plot when studies were grouped by categorical moderators (Automation Level, VTOR, ATOR, VITOR). The result of the meta-regression revealed only the VTORs have influenced the relation between NDRT and TOT, which can explain 7.59% of the heterogeneity (Q = 23.03, p = 0.006). Others did not account for the heterogeneity. Moreover, the meta-regression on the continuous moderators (Age, Sex Ratio, Driving Experience) received similar results. Age and Driving Experience had a negative relation with Hedge's g and the coefficient for sex ratio was positive. Driving experience explained 17.95% of the heterogeneity (Q = 10.36, p = 0.003), others had no contribution.

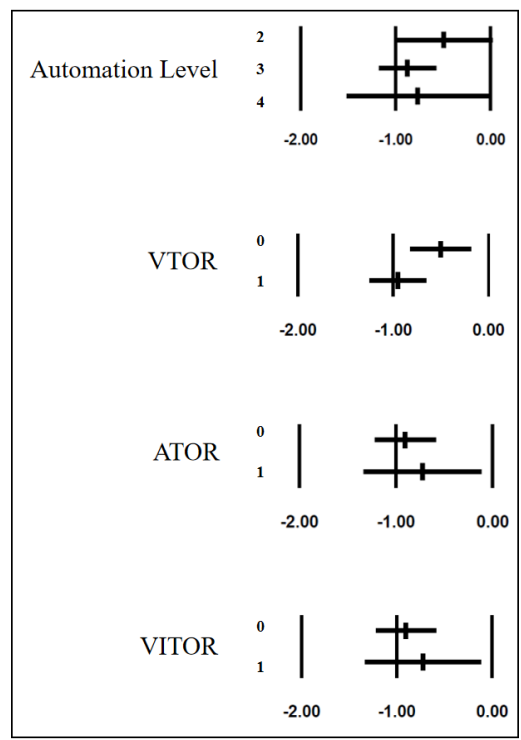

Figure 8. Categorical moderators for Vi Me Mo

3.4.2 Vi Me

As there was only one category for ATOR (ATOR = 1), it had been removed from the analysis. The forest tree of the categorical moderators (Figure 9) showed the results after grouping. The further meta-regression corrected that not all of them had influenced the relation between NDRT and TOT, only the Automation Level had explained 6% of the heterogeneity (Q = 7.86, p = 0.05). A regression analysis on continuous variables indicated the Driving Experience had contributed 10.83% of the heterogeneity (Q = 6.06, p = 0.05), others did not show the sign of moderation. All of them had a negative coefficient on Hedge's g.

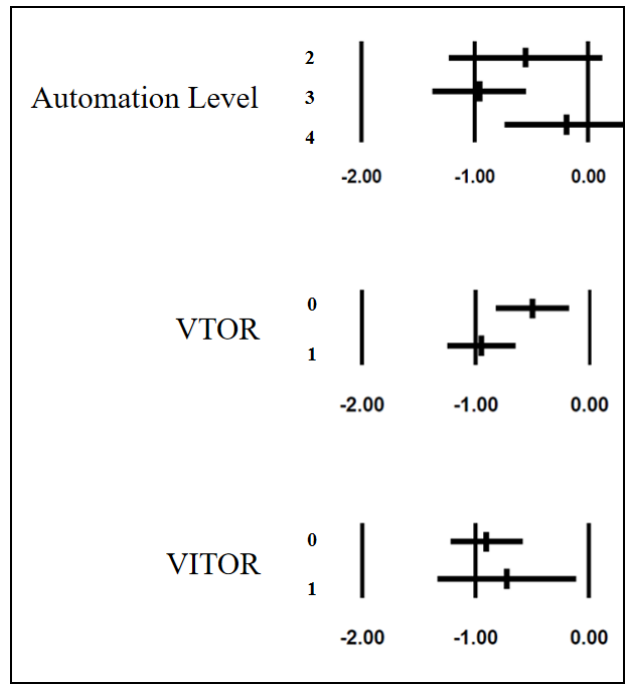

Figure 9. Categorical moderators for Vi Me

4. Discussion

4.1 Findings

This meta-analysis has included 37 studies related to NDRT and TOT in automated driving. In these studies, there were 52 experimental groups and 15 control groups. Nine types of NDRT were formed by five basic dimensions. The most commonly used NDRT was Vi Me Mo, which had 31 experimental groups in total. In another aspect, it proved that researchers preferred to use games, phone interactions, n-back tasks, and other sorting and tracking tasks as their secondary tasks.

The synthetical analysis examined all the experimental groups in detail. The results reflected several findings. Firstly, the mean takeover times of most studies, regardless of the NDRT type, fell in the range between 1000ms to 4500ms. Therefore, when facing a takeover request, drivers tended to take 1s to 4.5s to control the situation. Secondly, the Vi Me and the Vi Me Mo groups affected the takeover time the most, they also contained the largest number of studies. Indicating secondary tasks that require visual, mental, and motoric efforts may cause the longest takeover time.

The two-group analysis had a close view of the performance of Vi Me Mo and Vi Me groups. The results showed that both of them had a significant negative effect on the takeover time, which have confirmed the previous literature review and hypothesis that even though the automated system can help control the vehicle, before reaching fully automated driving, NDRTs still have a detrimental effect on drivers' performance (Louw & Merat, 2017). Comparing the results, this paper found that Vi Me Mo tasks had a larger effect than Vi Me tasks. Because motoric tasks would keep drivers' hands away from the steering wheel, this finding supported that hands-off driving could make drivers' performance worse than the hands-on situation.

The results of moderator analysis revealed several interesting findings. Firstly, for both Vi Me and Vi Me Mo groups, Age had a negative relation with Hedge's g, which means the same secondary task tended to have a greater effect on the takeover time on older drivers than younger drivers. Li et al. (2019) have found the same result. They suggested several reasons, like older drivers can be more cautious, older people have slower reactions, and a decline in psychomotor ability. Secondly, for both Vi Me and Vi Me Mo groups, Driving Experience had a negative relation with Hedge's g, which means people with longer driving experience tend to have a longer takeover time. In this paper, the driving experience was defined by the number of years driving non-automated vehicles. Morrison and Brantner (1992) suggested that similar but not exactly the same task experience can inhibit learning. In this case, it resulted in participants with longer driving experience may adapt to the automated vehicle slower than participants with shorter driving experience. Thus, the TOT became longer. People with longer driving experience may also be older, which was consistent with the second finding. Thirdly, the contradictory results of the relation between Sex Ratio and Hedge's g from Vi Me and Vi Me Mo did not conclude. Fourthly, for the automation level, level 2 showed abnormal better takeover time. However, there were only one and two studies in Vi Me and Vi Me Mo group, the result may not be valid. On the other hand, level 4 automation resulted in a smaller takeover time than level 3, with a relatively greater sample size. The level of automation was also a valid moderator of the effect size. This can be an affirmative finding because it may prove that higher automation level vehicles can be safer. Lastly, the presence of auditory and vibrotactile takeover requests can decrease the takeover time. In contrast, the Hedge's g became greater when VTOR was 1. Whereas, the sample size for category 0 was 2, which may be too small to draw any conclusion.

4.2 Limitations

One of the biggest limitations of this study was that the sample size was not large enough. Because of the lack of control groups, only Vi Me Mo and Vi Me could be conducted the two-group analysis. Five types of NDRT modality (Au, Mo, Au Me, Vi Mo, Vi Au Mo) had less than 5 sets of data. This limitation was caused by the practicability of experiments. In reality, the most common non-driving-related tasks are interaction with smartphones and laptops or using the in-car tablet to watch videos and play games. In order to restore the scene, Vi Me and Vi Me Mo tasks were the most frequently used NDRT. It was reasonable for other modalities to have a smaller sample size.

Another limitation was that some reviewed literature were lack of important information. When doing the meta-regression in the moderator analysis, variables like participants' age and the automation level of the apparatus were not declared in some articles. Those missing data were left blank and removed from the analysis.

Moreover, even though seven moderators were discussed in this paper, it is possible that there are unobserved but significant variables that can affect the takeover time.

Finally, when reviewing the eligible studies, one noticeable similarity among all the studies was that every study was using an indoor simulator to experiment. Because participants were informed that the experiment was not on-road, participants may reduce their alertness as their faults can cause no consequence compared to the real situation. (Carsten & Jamson 2011) Furthermore, researchers have found one main limitation of using VR autonomous vehicles, which was the simulation cannot perfectly demonstrate the real road condition. For instance, when simulating the train approach, the display of sound has flaws in detail that can destruct participants' immersive experience. (Zou et al., 2021)

5. Conclusion and Future Works

This meta-analysis investigated 37 studies on the drivers' takeover time when non-driving-related tasks were involved. NDRTs' negative effect on the takeover time has been proved. The correlations and contribution on the heterogeneity between moderators and the effect size were calculated. These findings may support the design and improvement of the ADS. The TOT can never be neglected, even though the automation level has reached the level 3 and 4. It is particularly important to customize the control system according to the characteristics of the driver.

In the future, this article suggested that studies could investigate the difference between simulators and real-life automated driving. Moreover, the modality of the tasks can be more diverse. There was a lack of data for many dimensions, like Au and Vi Au. For the engineering design, sensors and warning systems can be improved to adjust the time buffer to a more abundant level.

Acknowledgments

This research did not receive any specific grant from funding agencies in the public, commercial, or not-for-profit sectors.

Appendix

Table A. Documentation of all eligible papers

| Number | Title | Citation | NDRT Coding | Task | Study Name | Sample Size | Mean TOT (ms) | TOT Standard Deviation (ms) | Participants' average age |
|---|---|---|---|---|---|---|---|---|---|
| 1 | Takeover Time in Highly Automated Vehicles: Noncritical Transitions to and From Manual Control | Eriksson & Stanton, 2017 | N | N/A | Eriksson 2017 N | 26 | 6058 | 4863 | 32.37 |
| | | | Vi Me | Reading | Eriksson 2017 Vi Me | 26 | 6941 | 4329 | |
| 2 | Effects of non-driving-related tasks on takeover performance in different takeover situations in conditionally | Dogan, Honnêt, Masfrand, & Guillaume, 2019 | Vi Me Mo | Smartphone Interaction | Dogan 2019 Vi Me Mo | 21 | 5110 | 2500 | 38.00 |
| | | | Vi Au Me | Watch a Video | Dogan 2019 Vi Au Me | 21 | 5210 | 2470 | |
| | automated driving | | | | | | | | |
| 3 | Noncritical State Transitions During Conditionally Automated Driving on German Freeways: Effects of Non–Driving Related Tasks on Takeover Time and Takeover Quality | Naujoks, Purucker, Wiedemann & Marberge, 2019 | N | N/A | Naujoks 2019 N | 14 | 3180 | 1370 | |
| | | | Au | Audio book | Naujoks 2019 Au | 17 | 3480 | 1440 | 55.00 |
| | | | Vi Me | Reading | Naujoks 2019 Vi Me | 33 | 5459 | 1992 | |
| | | | Vi Me Mo | Tablet game | Naujoks 2019 Vi Me Mo | 16 | 4010 | 1740 | |
| 4 | Non-driving-related tasks, workload, and takeover performance in highly automated driving contexts | Yoon & Ji, 2019 | Vi Me Mo | Tablet game, smartphone interaction | Yoon 2019 Vi Me Mo | 54 | 1610 | 720 | 29.10 |
| | | | Vi Au Me | Watch a video | Yoon 2019 Vi Au Me | 27 | 1800 | 580 | 32.32 |
| 5 | Take-Over Performance and Safety Analysis Under | Wu, Wu, Lyu, & Zheng, 2019 | Vi Me Mo | Tablet game, n-back task | Wu 2019 Vi Me Mo | 56 | 2800 | 1192 | 39.00 |

| | | | | | | | | |
|---|---|---|---|---|---|---|---|---|
| | Different Scenarios and Secondary Tasks in Conditionally Automated Driving | | | | | | | |
| 6 | Transition of control in a partially automated vehicle: Effects of anticipation and non-driving-related task involvement | Dogan et al., 2017 | N | N/A | Dogan 2017 N | 14 | 2050 | 770 | 39.00 |
| | | | Vi Me Mo | Smartphone interaction | Dogan 2017 Vi Me Mo | 14 | 2500 | 700 | |
| 7 | Coming back into the loop: Drivers' perceptual-motor performance in critical events after automated driving | Louw et al., 2017 | N | N/A | Louw 2017 N | 15 | 3700 | 1220 | 39.00 |
| | | | Vi Me Mo | N-back task | Louw 2017 Vi Me Mo | 15 | 4110 | 1070 | |
| 8 | Driver compliance to | Forster, | N | N/A | Forster 2017 | 17 | 7840 | 3160 | 29.00 |

| | | | | | N | | | | |
|---|---|---|---|---|---|---|---|---|---|
| | take-over requests with different auditory outputs in conditional automation | Naujoks, Neukum, & Huestegge, 2017 | Vi Me | Reading | Forster 2017 Vi Me | 17 | 9120 | 550 | |
| 9 | The Sooner the Better: Drivers' Reactions to Two-Step Take-Over Requests in Highly Automated Driving | Epple, Roche, & Brandenburg, 2018 | Vi Me | Read-question-and-give-answer task | Epple 2018 Vi Me | 40 | 3990 | 1740 | 27.00 |
| | | | N | N/A | Yoon2 2019 N | 19 | 2040 | 1014 | |
| 10 | The effects of takeover request modalities on highly automated car control transitions | Yoon, Kim, & Ji, 2019 | Au Me | Phone conversation | Yoon2 2019 Au Me | 11 | 2110 | 1079 | 30.10 |
| | | | Vi Me Mo | Smartphone interaction | Yoon2 2019 Vi Me Mo | 17 | 2410 | 1131 | |
| | | | Vi Au Me | Watch a video | Yoon2 2019 Vi Au Me | 19 | 1970 | 948 | |

| | | | | | | | | |
|---|---|---|---|---|---|---|---|---|
| 11 | Take-over again: Investigating multimodal and directional TORs to get the driver back into the loop | Petermeijer, Bazilinskyy, Bengler, & De Winter, 2017 | Vi Me Mo | SuRTs | Petermeijer 2017 Vi Me Mo | 24 | 1570 | 380 | 27.90 |
| 12 | Directional tactile alerts for take-over requests in highly automated driving | Cohen-Lazry, Katzman, Borowsky, & Oron-Gilad, 2019 | Vi Me Mo | Tablet game | Cohen-Lazry 2019 Vi Me Mo | 8 | 1120 | 340 | 24.90 |
| 13 | Investigation of older driver's takeover performance in highly automated vehicles in adverse weather conditions | Li, Blythe, Guo, & Namdeo, 2018 | Vi Me | Reading | Li 2018 Vi Me | 37 | 3610 | 1790 | 26.05 |
| 14 | An HMI Concept to | Yang, | Vi Me | SuRTs | Yang 2018 Vi | 25 | 3210 | 1010 | 24.96 |

| | | | | | | | | |
|---|---|---|---|---|---|---|---|---|
| | Improve Driver's Visual Behavior and Situation Awareness in Automated Vehicle | Karakaya, Dominioni, Kawabe, & Bengler, 2018 | Mo | | Me Mo | | | |
| 15 | Transitioning to manual driving requires additional time after automation deactivation | Vogelpohl, Kühn, Hummel, Gehlert, & Vollrath, 2018 | N | N/A | Vogelpohl 2018 N | 60 | 2370 | 1190 |
| | | | Vi Me | Reading | Vogelpohl 2018 Vi Me | 60 | 4590 | 1970 | 36 |
| | | | Vi Me Mo | Tablet game | Vogelpohl 2018 Vi Me Mo | 60 | 5070 | 1940 |
| 16 | Let Me Finish before I Take Over: Wintersberger Towards Attention Aware Device Integration in Highly | Wintersberger, Riener, Schartmüller, Frison, & Weigl, 2018 | Vi Me Mo | Smartphone interaction | Wintersberger 2018 Vi Me Mo | 18 | 2590 | 740 | 25.71 |
| | Automated Vehicles | | | | | | | | |
| 17 | Investigating the effects of age and disengagement in driving on driver's takeover control performance in highly automated vehicles | Li, Blythe, Guo, & Namdeo, 2019 | Vi Me Mo | N-back task | Li 2019 Vi Me Mo | 76 | 3790 | 1470 | 49.21 |
| | | | N | N/A | Li 2019 N | 76 | 2460 | 870 | |
| 18 | The comparison of auditory, tactile, and multimodal warnings for the effective communication of unexpected events during an automated driving scenario | Geitner, Biondi, Skrypchuk, Jennings, & Birrell, 2019 | Vi Au Mo | Dictation task | Geitner 2019 Vi Au Mo | 45 | 750 | 8 | |
| 19 | The Effects of | Wan & Wu, | N | N/A | Wan 2017 N | 36 | 2600 | 1140 | 25.40 |
| | Vibration Patterns of Take-Over Request and Non-Driving Tasks on Taking-Over Control of Automated Vehicles | 2018 | Vi Me | Reading | Wan 2017 Vi Me | 36 | 3400 | 1210 | |
| | | | Vi Me Mo | Smartphone interaction | Wan 2017 Vi Me Mo | 36 | 3140 | 1100 | |
| | | | Vi Au Me | Watch a video | Wan 2017 Vi Au Me | 72 | 3280 | 6900 | |
| 20 | Workaholistic: on balancing typing- and handover-performance in automated driving | Schartmüller, Riener, Wintersberger, & Frison, 2018 | N | N/A | Schartmuller 2018 N | 21 | 1510 | 570 | 23.10 |
| | | | Vi Me Mo | Smartphone interaction | Schartmuller 2018 Vi Me Mo | 21 | 1910 | 440 | |
| 21 | Impact of training and in-vehicle task performance on manual control recovery in an automated car | Payre, Cestac, Dang, Vienne, & Delhomme, 2017 | N | N/A | Payre 2017 N | 113 | 4140 | 1780 | 40.30 |
| | | | Vi Me Mo | Anagrams and labyrinth game | Payre 2017 Vi Me Mo | 113 | 5020 | 1590 | |
| 22 | The impact of auditory continual feedback on take-overs in Level 3 automated vehicles | Cohen-Lazry, Borowsky, & Oron-Gilad, 2020 | Vi Me Mo | Tablet game | Cohen-Lazrt 2020 Vi Me Mo | 18 | 2730 | 920 | 25.16 |
| 23 | Tactical-Level Input with Multimodal Feedback for Unscheduled Takeover Situations in Human-Centered Automated Vehicles | Manawadu et al., 2018 | Vi Me Mo | N-back task | Manawadu 2018 Vi Me Mo | 11 | 6270 | 190 | 28.60 |
| 24 | Effects of secondary tasks and display position on glance behavior during partially automated driving | Hensch et al., 2020 | Vi Me | Reading | Hensch 2020 Vi Me | 50 | 7240 | 9690 | 37.90 |
| | | | Vi Me Mo | SuRTs | Hensch 2020 Vi Me Mo | 50 | 4760 | 5420 | |
| 25 | Multimodal warning | Yun & Yang, | Vi Me | Smartphone interaction | Yun 2020 Vi | 41 | 2320 | 740 | 26.20 |
| | design for take-over request in conditionally automated driving | 2020 | Mo | | Me Mo | | | | |
| 26 | Task Interruption and Control Recovery Strategies After Take-Over Requests Emphasize Need for Measures of Situation Awareness | Vogelpohl, Gehlmann, & Vollrath, 2020 | N | N/A | Vogelpohl 2020 N | 47 | 1892 | 815 | |
| | | | Vi Me Mo | Sorting and tracking task | Vogelpohl 2020 Vi Me Mo | 94 | 4566 | 3813 | 22.60 |
| 27 | Should the Urgency of Visual-Tactile Takeover Requests Match the Criticality of Takeover Situations? | Roche & Brandenburg, 2020 | Vi Me Mo | Question and answer game | Roche 2020 Vi Me Mo | 52 | 1800 | 620 | 27.20 |
| 28 | The relationship | Rauffet, | Vi Me | Tablet game | Rauffet 2020 | 28 | 4790 | 2040 | 46.00 |
| | between level of engagement in a non-driving task and driver response time when taking control of an automated vehicle | Botzer, Chauvin, Saïd, & Tordet, 2020 | Mo | | Vi Me Mo | | | | |
| 29 | Design concept for a visual, vibrotactile and acoustic take-over request in a conditional automated vehicle during non-driving-related tasks | Mueller, Ogrizek, Bier, & Abendroth, 2018 | Vi Me Mo | N-back task | Mueller 2018 Vi Me Mo | 21 | 1170 | 200 | 27.30 |
| 30 | User Experience with Increasing Levels of | Reilhac, Hottelart, | Vi Me Mo | Smartphone interaction | Reihac 2017 Vi Me Mo | 42 | 2480 | 566 | 36.50 |

| | Vehicle Automation: Overview of the Challenges and Opportunities as Vehicles Progress from Partial to High Automation | Diederichs, & Nowakowski, 2017 | N | N/A | Reihac 2017 N | 42 | 1860 | 361 | |
| 31 | Text Comprehension: Heads-Up vs. Auditory Displays | Schartmüller, Weigl, Wintersberger, Riener, & Steinhauser, 2019 | Vi Me Mo | Smartphone interaction | Schartmuller 2019 Vi Me Mo | 32 | 4870 | 1070 | 32.37 |
| 32 | Voices in Self-Driving Cars Should be Assertive to More Quickly Grab a Distracted Driver's | Wong, Brumby,Babu, & Kobayashi, 2019 | Vi Me Mo | Tablet game | Wong 2019 Vi Me Mo | 20 | 9540 | 83 | 26.30 |

| | | | | | | | | |
|---|---|---|---|---|---|---|---|---|
| | Attention | | | | | | | |
| 33 | Assessment and Support of Error Recognition in Automated Driving | Spießl, 2011 | N | N/A | Spiebl 2011 N | 29 | 1360 | 170 | |
| | | | Vi Me Mo | Sorting and tracking task | Spiebl 2011 Vi Me Mo | 29 | 1660 | 300 | 29.47 |
| 34 | Effect of Levels of Automation and Vehicle Control Format on Driver Performance and Attention Allocation | Deng 2017 | Vi Mo | Mimic CD-player | Deng 2017 Vi Mo | 11 | 761 | 229 | 25.00 |
| 35 | Smart S3D TOR: Intelligent Warnings On Large Stereoscopic 3D Dashboards During Take-Overs | Weidner & Broll, 2019 | Vi Me Mo | Smartphone interaction | Weidner 2019 Vi Me Mo | 52 | 1748 | 285 | 31.90 |
| 36 | Manual takeover after highly automated | Dogan, Yousfi, Bellet, | Vi Me | Mah-Jong game | Dogan 2021 | 62 | 2540 | 510 | 32.00 |

| | | | | | | | | |
|---|---|---|---|---|---|---|---|---|
| | driving | Tijus, & Guillaume, 2021 | Mo | | Vi Me Mo | | | |
| | | | N | N/A | Yoon 2021 N | 29 | 2256 | 750 | 32.00 |
| | | | Au Me | Conversing with a passenger, Cellphone talk | Yoon 2021 Au Me | 60 | 2521 | 722 | 30.90 |
| 37 | Modeling takeover time based on non-driving-related task attributes in highly automated driving | Yoon, Lee, & Ji, 2021 | Au | Listening to music | Yoon 2021 Au | 29 | 2172 | 560 | 30.90 |
| | | | Vi Au Me | Video watching | Yoon 2021 Vi Au Me | 30 | 2369 | 450 | 30.90 |
| | | | Vi Me | Reading | Yoon 2021 Vi Me | 30 | 2702 | 750 | 30.90 |
| | | | Vi Me Mo | Texting, Surfing, Smartphone game | Yoon 2021 Vi Me Mo | 89 | 2661 | 788 | 30.90 |
| | | | Mo | Holding a drink | Yoon 2021 Mo | 30 | 2745 | 1090 | 30.90 |